\documentclass[aps,prb,groupedaddress,twocolumn,10pt]{revtex4-1}

\begin{document}

\newcommand{\be}{\begin{equation}}
\newcommand{\ee}{\end{equation}}
\newcommand{\bea}{\begin{eqnarray}}
\newcommand{\eea}{\end{eqnarray}}

\title{Connecting the SYK dots}

\author{D. V. Khveshchenko}
\affiliation{Department of Physics and Astronomy, University of North Carolina, Chapel Hill, NC 27599}

\begin{abstract}
\noindent
We study a putative (strange) metal-to-insulator transition in a granular array of the Sachdev-Ye-Kitaev (SYK) quantum dots, each occupied by a large number $N\gg 1$ of charge-carrying fermions. 
Extending the previous studies we complement the SYK couplings by the physically relevant Coulomb interactions and  
focus on the effects of charge fluctuations, evaluating the conductivity and density of states. The latter were found to demonstrate marked changes of behavior when the effective inter-site tunneling becomes comparable to the renormalized Coulomb energy, {thus signifying the transition in question}. 
\noindent
\end{abstract}
\maketitle

\vspace{0.3cm}
\noindent
{\bf {\it Introduction}} 
\vspace{0.3cm}

The recent upsurge of interest in the 
SYK and related models of (super)strongly interacting dispersionless fermions with all-to-all $q$-body couplings has been driven, among other things, by the hopes of utilizing them as (asymptotically) solvable examples of the so-called non-Fermi liquid (NFL) behavior. 

Originally, the SYK reincarnation [1-3] of the parent SY (Sachdev-Ye) [4-8] model was formulated in terms of neutral Majorana fermions that would be abundantly present in the recent theoretical constructs (albeit less so in their  
attempted experimental realizations). 
However, in order to account for the physically relevant 
charge (alongside energy) dynamics one needs to use charged (complex or Dirac) fermions [9-10].

Regardless of the nature of its constituent fermions, though, the original SYK model lacks any spatial dispersion and, therefore, can be best thought of as a (zero-dimensional) 'quantum dot'. As such, this system exhibits a characteristic 'local NFL' behavior characterized by the anomalous power-law decay of its temporal (but not spatial) correlations [1-8]. 

Therefore, while predicting some markedly novel features [11-15] in mesoscopic charge and heat transport through 
{ its proposed (but not yet implemented) realizations in the irregularly shaped graphene flakes, multi-connected Quantum Hall setups, semiconductor wires and quantum dots, as well as topological insulator surfaces [16-20]}, the complex SYK model still needs to to be extended into the spatial dimensions before applying it to the analysis of any documented higher-dimensional NFL system. 

In the early 'SYK-lattice' constructions [21-36], the individual SYK dots would be arranged in a regular array by adding short-range (nearest-neighbor) one- and/or two-body entangling terms into the Hamiltonian. 
Alternatively, the immobile SYK  
fermions would be hybridized with their conduction counterparts  
or subject to long-range and distance-dependent many-body couplings.

Such generalizations allow for a variety of the NFL regimes, some of which are even capable of ostensibly reproducing, e.g., the ubiquitous linear temperature dependence of electrical resistivity [37-40]. 

Still different is a formulation of the SYK model directly in the momentum space which approach appears to be miraculously 
successful in providing nothing short of a quantitative agreement between the computed linear resistivity and its measured values in a sizeable number of the well studied experimental compounds [41]. 

Among the central issues studied { in the context of the SYK-lattices} are putative phase transitions between the parent NFL SYK state 
({ often referred to as} 'strange metal' (SM)) and a more conventional (disordered) Fermi liquid (FL) or, alternatively, a (many-body) Mott insulator (MI). 

{ However, the previous analyses were, by and large, limited to the effects of the 
(somewhat exotic) SYK-type entangling correlations, thus leaving out the far more mundane (yet, physically relevant and practically unavoidable) charge couplings of the Coulomb origin that are going to affect any feasible solid-state  implementation of the SYK system, including those of Refs.[16-20].

More specifically, in such electron-based setups the $SYK_4$ interaction itself would be simulated through the geometrically randomized intra-site Coulomb couplings.  That alone makes it anything but consistent to neglect the (non-random) inter-site charging effects, if a viable SYK-lattice were to be engineered out of the single-site SYK building blocks
akin to those proposed in Refs.[16-20].  

In defence of the earlier studies of Refs.[21-36], any concrete practical realizations of the SYK-lattices did not seem to be particularly high on their agenda, while speculations regarding their potential applications to such long-standing  experimental challenges as the high-$T_c$ cuprates or heavy-fermion materials [37-39, 41] would be made 
largely as a matter of custom. 
   
In the present note, we fill in the gap by investigating the charging effects in a manner similar to that utilized in the context of the ordinary (FL) granular electronic materials [42-49].} 

\vspace{0.3cm}
\noindent
{\bf {\it SYK model of charged fermions}} 
\vspace{0.3cm}

The Hamiltonian of the SYK array can be 
written in terms of the complex fermions $\psi_{i\alpha}$ localized at the dot $i$ and carrying a flavor $\alpha=1,\dots, q$ 
(hereafter $q$ is an even integer) 
\be
H=H_{SYK}+H_T+H_C
\ee
which sum includes the customary SYK intra-site 
$q$-fermion couplings and a chemical potential $\mu$ 
\be
H_{SYK}=\sum_{i;\alpha,\dots\beta}
{J_i^{\alpha\dots\beta}}
\psi^{\dagger}_{i\alpha}
\dots
\psi_{i\beta}
-
\mu
\psi^{\dagger}_{i\alpha}
\psi_{i\alpha}
\ee
as well as inter-site tunneling 
\be
H_T=
\sum_{ij;\alpha\beta}t^{\alpha\beta}_{ij}\psi^{\dagger}_{i\alpha}\psi_{j\beta}
\ee
and, both, intra- and inter-site charging energies
\be
H_C=\sum_{i,j;\alpha,\beta}
{U_{ij}\over 2}
(\psi^{\dagger}_{i\alpha}\psi_{i\alpha}-Q_{i})
%\nonumber\\
(\psi^{\dagger}_{j\beta}\psi_{j\beta}-Q_{j})
\ee
which include the offset charges $Q_{i}$, if any (in units of electron charge). 

In the SYK model the $q$-fermion amplitudes 
$J^{\alpha\dots\beta}_{i}$ in (2) are treated as Gaussian random variables with the time- and state-independent variances 
\be 
<J_{i}^{\alpha\dots \beta}J_{j}^{\alpha^{\prime}\dots\beta^{\prime}}>
={J^2\over N^{q-1}}
\delta_{ij}\delta^{\alpha\alpha^{\prime}}\dots\delta^{\beta\beta^{\prime}} 
\ee
Averaging (1) over such distribution results in introducing temporally bi-local 
$2q$-fermion terms to the effective action
[1-8].

Likewise, in most of the previous studies of the SYK-lattices [21-36] the tunneling amplitudes in (3) would be treated as random, with the dispersion 
\be 
<t^{\alpha\beta}_{ij}
t^{\alpha^{\prime}\beta^{\prime}}_{ij}>=
{t^2\over N}\delta_{j,i+{\hat \mu}}\delta^{\alpha\alpha^{\prime}}
\delta^{\beta\beta^{\prime}} 
\ee
where $\hat \mu$ is one of $z$ (coordination number) prime vectors of the SYK lattice. Upon averaging, the tunneling term would then result in the inter-site $SYK_2$-type coupling. 

Alternatively one might view such amplitudes as fixed at some $N$-independent value and diagonal in the 'flavor' space, 
$t^{\alpha\beta}_{ij}=t\delta_{ij}\delta^{\alpha\beta}$. 

Lastly, the offset charges can also be thought of random variables, the degree of their disorder ranging from strong (described by a uniform distribution within the entire interval $-1/2<Q_i<1/2$, which situation 
might be appropriate for naturally assembled networks) to weak (confining the charges to certain values such as, e.g.,
$|Q_i|\ll 1$, more suitable for artificially patterned arrays). 

A systematic investigation into 
those different situations would definitely be warranted. However, in much of what follows we drop the offset charges altogether, focusing on the regimes that are farthest from (accidental) degeneracies. {In that regard, our main goal will be to demonstrate that a conducting state could emerge even under the least conducive conditions.

To that end, the previous studies of the hybrid model with the intra-site $SYK_4$
and inter-site $SYK_2$ couplings - but without any charging effects - have repeatedly reported observing} a crossover from the SM described by the ergodic $SYK_4$ model to a disordered FL state corresponding to the non-ergodic $SYK_2$ one at temperatures of order the effective fermion kinetic energy
$t^*\sim t^2/J$ [21-36].
{However, a potentially critical impact of the Coulomb blockade (CB) due to the charging energy (4) has not been investigated. 

By contrast, in the conventional (FL) granular arrays the latter has long been known to invariably drive the system insulating [42-49]. 
On the other hand, a coupling to some dissipative sub-Ohmic 
bath was shown to quell the CB, thereby promoting a conducting state [42-45].
 
Below, we demonstrate that in the problem at hand the role of such sub-Ohmic bath is played by the intra-site SYK correlations themselves, thus enabling the metal-to-insulator transition (MIT) in the granular SYK systems even in the presence of the charging effects.}  

\vspace{0.3cm}
\noindent
{\bf {\it SYK strange metal}} 
\vspace{0.3cm}

First, we consider the $U\to 0$ limit 
where the influence of the tunneling term 
on the on-site fermion propagator 
\be
G_{i}(\tau)={1\over N}\sum_{\alpha}<\psi_{i\alpha}(\tau)\psi^{\dagger}_{i\alpha}(0)>=
(\partial_{\tau}-\mu-\Sigma_{i})^{-1}
\ee
is captured in terms of the intra-site 
self-energy 
\be
\Sigma_{i}(\tau)={J^2}G^{q-1}_{i}(\tau)+\sum_jt_{ij}G_{j}(\tau)t_{ji}
\ee
where the first term represents the 
effect of the SYK correlations. 
This approximation can be further improved, thereby systematically recovering all the (even order) tunneling processes.

For $t,U\to 0$ and $N\gg 1$ the fermion propagator (7) takes the spatially local
SYK form 
\be
G_{ij}(\tau)=\delta_{ij}G_{i}(\tau)
=
\delta_{ij} 
A sgn(\tau)
e^{sgn(\tau) \pi{\cal E}}/(J\tau)^{2\Delta}
\ee
where $\Delta=1/q$ while the prefactor  
$A$ is a known function of $q$ and 
the dimensionless parameter ${\cal E}(\mu)$ controls the fermion density [1-3]. 
 
Apart from the mean-field solution (9), in the no-tunneling/zero-charging energy limit the theory (1) possesses a manifold of nearly degenerate solutions which are continuously connected to (9) by virtue of 
arbitrary diffeomorphisms of the thermodynamic time 
variable $\tau\to f_{i}(\tau)$,   
obeying the boundary conditions $f_{i}(\tau+1/T)=f_{i}(\tau)+1/T$, combined with local $U(1)$ phase rotations [1-3, 9-10]
\be
G_{i}(\tau_1,\tau_2)
=A
e^{i\Phi_{i}(\tau_1)-i\Phi_{i}(\tau_2)}
{\large (}
{\partial_{\tau}f_{i}(\tau_1)\partial_{\tau}f_{i}(\tau_2)\over (f_{i}(\tau_1)-f_{i}(\tau_2))^2}{\large)}^{\Delta}
\ee 
In particular, a finite-temperature counterpart of (9) can be obtained by the conformal mapping onto a thermal circle, $\tau\to\sin(\pi T\tau)/\pi T$.

In addition to being spontaneously broken by the particular choice of the mean-field solution (9) down to the subgroup formed by the Mobius transformations  $SL(2,R)$, the reparametrization symmetry gets violated 
explicitly by the temporal gradients 
$\partial_{\tau}f$, as well as the 
tunneling and Coulomb terms in Eq.(1). 

Importantly, the $U(1)$ phase fluctuations have no effect on the intra-site SYK terms while the inter-site tunneling terms can be heavily impacted. 

As shown in the earlier studies of the SYK-lattices, the low-energy collective charge and energy fluctuations about the mean-field solution (9) 
can propagate even in the absence of a bare single-particle dispersion ($<t_{ij}>=0$), 
as manifested by the same-site localization of the fermion propagator (9). 

Small fluctuations are governed by the Gaussian action 
\bea
S_{G}({\delta\Phi},\delta f)=
\sum_{\bf q}
\int_{\omega}
({1\over 2E_C}|{\delta\Phi}|^2
|\omega|(|\omega|+{\cal D}{\bf q}^2)
\nonumber\\
+{\gamma N\over 2J}
|{\delta f}|^2
|\omega|(|\omega|+{\cal D}^{\prime}{\bf q}^2)
(\omega^2-(2\pi T)^2))
\eea 
where $\gamma$ is a $q$-dependent coefficient vanishing for $q=2$ [9-10] and  
the momentum sum goes over 
the Brillouin zone of the SYK lattice. 

The diffusion coefficients  
$
{\cal D}
$
and
$ 
{\cal D}^{\prime}
$
pertain to the spatial spreading of charge and energy, respectively. 
Their values are expected to comply with the lower bound of order $(t^*a)^2/T$ (here $a$ is the 
lattice constant) in the high-$T$ regime
where the inelastic SYK scattering becomes the fastest equilibration mechanism [9-10]. 

 The phase fluctuations 
${\delta\Phi}$ described by the first term in (11) develop below the (independent of $N$) charging energy $E_C$ which, alongside the intra-/inter-site capacitive couplings, includes the 
energy of induced voltages, 
$E^{-1}_C=U^{-1}+(\partial Q/\partial\mu)_T$,
the second term being due to the 
fermion compressibility.  

In turn, the second term in (11) 
describes the low-energy dynamics of the SYK reparametrization mode and originates from the intra-site Schwarzian derivative 
$Sch{\{}\tan{\pi Tf},\tau {\}}$ defined as follows: 
$
Sch{\{}y,x {\}}=
({y^{\prime\prime\prime}/y^{\prime}})-
({3y^{\prime\prime}/2y^{\prime}})^2
$ [1].

Upon the change of variables 
$\partial_{\tau}f_{i}=e^{\phi_{i}}$ it yields,
in addition to the quadratic term in Eq.(11),
the non-Gaussian ('Liouville') interaction,
$
S_{NG}({\delta f})=(2\pi T)^2
{\gamma N\over J}
\sum_i\int_{\tau} e^{2\phi_{i}(\tau)}
$
[1-3].

Importantly, the $\phi$ fluctuations can only be activated at exceedingly low energies/temperatures $\omega, T\lesssim J/N$ while 
above that scale their effect can be neglected.

Whenever present such fluctuations  provide 'gravitational dressing' of any products of the vertex operators $e^{\phi_{i}(\tau)}$. This effect can be elegantly taken into account by making use of the exactly solvable Liouville quantum mechanics deformed by the 'quench' potential acting between the consecutive applications of such operators [50-51].  

As the result, an arbitrary power $p$ of the fermion propagator 
of an isolated SYK system  
develops a universal asymptotic behavior
for all the integer $p$ and $q>2$ 
\be
<G^p_{i}(\tau)>\sim N^{3/2-2\Delta p}/(J\tau)^{3/2}  
\ee,
where the averaging stands for a functional integration over the soft 'Schwarzian' modes $f_i(\tau)$ [50-51].

Moreover, if the local reparametrizations were locked into one global transformation $f(\tau)$, thus drastically reducing the space of the low-energy deformations of the solution (9), then the universal asymptotic (12) would even be shared by the multi-local products  $<\prod_iG_i^{p_i}(\tau)>$. 

\vspace{0.3cm}
\noindent
{\bf {\it Phase fluctuations}} 
\vspace{0.3cm}

\noindent
The phenomenological action (11) conceived in Ref.[9-10] under the customary assumption of a regular gradient expansion does not account for any singular (temporally non-local) effects of the SYK correlations. Nor, does it allow for a systematic derivation of any non-Gaussian terms. 

The classic studies of such effects in the conventional (FL) granular materials were facilitated by representing the fermion operator as a product of its energy- and charge-related constituents, 
$
\psi_{i\alpha}=\chi_{i\alpha}e^{i\Phi_i}
$ [42-45].

Correspondingly, the fermion propagator factorizes 
\bea
G_{ij}(\tau)={\cal G}_{ij}(\tau)D_{ij}(\tau)=\nonumber\\
<\chi_{i}(\tau)\chi^{\dagger}_{j}(0)>
<e^{i\Phi_i(\tau)}e^{-i\Phi_i(0)}>
\eea
onto its 'energy' and 'charge' components.

The 'fractionalized' fermionic degrees of freedom $\chi_{i\alpha}$ can still be traded for the SYK field $\phi$ corresponding to the quasiparticle-hole excitations, while the phase variable $\Phi$ describes the collective ('plasmon') mode. 

As already mentioned, at a sizeable charging energy the phase fluctuations dominate
in the entire range $J/N<T<E_C$
where the SYK fluctuations remain frozen.

Besides affecting the $\cal G$ propagator, as per Eqs.(7,8), the tunneling term (3)
introduces a (singular) non-Gaussian term into the effective action for the phase field 
\be
S_{NG}(\Phi)=
{1\over 2}\sum_{ij}\int_{\tau_1,\tau_2} 
K_{ij}
(\tau_1-\tau_2)\cos(\Phi_{ij}(\tau_1)-\Phi_{ij}(\tau_2))
\ee
where 
$
\Phi_{ij}(\tau)=\Phi_{i}(\tau)-
\Phi_{j}(\tau)
$ and the trigonometric functional dependence 
stems from the intrinsic 
compactness of the phase 
variable subject to the periodic 
boundary condition, $\Phi_i(\tau+1/T)=\Phi_i(\tau)+2\pi n_i$.

The kernel 
$K_{ij}(\tau)=
t^2 {\cal G}_i(\tau){\cal G}_j(-\tau)
$
in the 'influence functional' (14)
represents the effect of a dissipative 
particle-hole bath on the phase dynamics. 

On the metallic side of the putative metal-insulator transition and for $T=0$ 
this kernel decays algebraically, albeit with different exponents
depending on whether or not the system is near criticality.
 
Deep in the FL phase and away 
from the critical regime 
the phase propagator $D_{ij}(\tau)$ remains nearly constant and the kernel reads 
\be
K_{ij}(\tau)=\delta_{j,i+{\hat \mu}}({gE_C^{2\epsilon}/\tau^{2-2\epsilon}})
\ee
where the strength of tunneling is 
quantified in terms of the dimensionless 'çonductance' 
$
g\sim t^2/J^{2-2\epsilon}E_C^{2\epsilon}
$. 

The time dependence is controlled by the exponent  
$
\epsilon=1-2\Delta
$ 
which varies between $0$ (FL, $q=2$) and $1$ ({ free dispersionless fermions}, $q\to\infty$), thereby making the kernal (15) generically sub-Ohmic for all $q>2$. 

This should be contrasted against the case of
an ordinary (FL) granular system     
where such a regime could only be attained in the presence of a sufficiently strong excitonic enhancement. Otherwise, the kernel (15) turns super-Ohmic due to the competing effect of orthogonality catastrophe [42-45]. 

At a would-be quantum critical point the system is expected to undergo a transition from the disordered ($<\cos\Phi_i>=0$, conceivably for $g<g_c$) insulating state governed by the Coulomb blockade (CB) to a dissipation-driven ordered ($<\cos\Phi_i>\neq 0$) conducting one for $g>g_c$. 
In the latter state, a condensation of the phase field, 
$D_{ij}(\tau\to\infty)=const$, implies 
a vanishing effective charging energy $E_C^{*}$. 

\vspace{0.3cm}
\noindent
{\bf {\it Mean-field analysis}} 
\vspace{0.3cm}

\noindent
In the critical regime, the system of coupled equations 
for the $\cal G$ and $D$ propagators reads
\bea
J^2\int_{\tau}[{\cal G}]^{q-1}_{ik}
(\tau_1-\tau){\cal G}_{kj}(\tau-\tau_2)+
\nonumber\\
t^2\int_{\tau} [{\cal G}D^2]_{ik}
(\tau_1-\tau){\cal G}_{kj}(\tau-\tau_2)=\delta_{ij}\delta(\tau_1-\tau_2)
\nonumber\\
t^2\int_{\tau}[{\cal G}^2D]_{ik}(\tau_1-\tau)D_{kj}
(\tau-\tau_2)=\delta_{ij}\delta(\tau_1-\tau_2)
\eea
Incidentally, similar equations and their solutions 
have been explored in a number of recent works 
dealing with the { mathematically related problem of the} transitions between metallic spin-glass and disordered 
FL states in the randomized 
Hubbard and $t-J$ models [52-55]. 

At the critical point, the spatially local and temporally algebraic behavior inherited from the pure SYK model  
extends all the way down to the lowest energies/temperatures.
In particular, the fermion propagator $\cal G$ retains its SYK behavior (4) 
with the fermion dimension $\Delta$ while the algebraically decaying phase correlator
\be 
D_{ii}(\tau)=B/(E_C\tau)^{2\Delta_{\Phi}}
\ee 
manifests the exponent 
$
\Delta_{\Phi}=\epsilon/2
$.

The dimensionless 
amplitudes $A$ and $B$ then satisfy the equations
\be
A^q+ \alpha{g}B^2A^2=1
~~~~~~~~
\beta{g}A^2B^2=1
\ee
which allow for a non-trivial solution provided that the numerical prefactors obey the condition 
$\alpha<\beta$.

Notably, the overall exponent governing the decay of the physical fermion propagator $G$ 
attains the FL value, $2\Delta_{\Phi}+2\Delta=1$, 
thereby connecting smoothly with that in the FL phase for $g>g_c$.

Thus, invoking 
the phase fluctuations appears to be instrumental for reconciling the seemingly conflicting predictions for the 
fermion dimension $[\psi]$ that one would obtain by approaching the quantum
critical point from the FL phase (where $[\psi]_{FL}=1/2$) by lowering $g$ towards $g_c$ at $T=0$, as compared to  
lowering $T$ within the SYK phase (where $[\psi]_{SYK}=\Delta$) at $g=g_c$.

The properties of the critical point 
can be further discerned by 
employing a mean-field analysis akin to those 
of Refs.[42-45]. To that end,
a two-component $O(2)$ bosonic variable $w_{1,2}=(\cos\Phi,\sin\Phi)$ (or, equivalently, one unimodular complex-valued variable $w=w_1+iw_2=e^{i\Phi}$) is promoted to a multi-component vector $w_{1,\dots,M}$ transforming under $O(M)$ and described 
by the 'dissipative non-linear $\sigma$-model' 
\bea
S_{NL\sigma}({\bf w},\lambda)=
\sum_i
\int_{\tau}
({1\over 2E_C}
(\partial_{\tau}{\bf w})^2
+
i\lambda({\bf w}^2-1))
\nonumber\\
+
{1\over 2}\sum_{ij}\int_{\tau_1,\tau_2} 
{K}^{\prime}_{ij}
(\tau_1-\tau_2){\bf w}_{i}(\tau_1){\bf w}_{j}(\tau_2)
~~~~~~~~~~~~\eea 
where the self-consistently determined near-critical kernel differs from (15)
due to the etxra $\Delta_{\Phi}$ 
\be
{K}^{\prime}_{ij}(\tau)=K_{ij}(\tau)
D_{ij}(-\tau)=
\delta_{j,i+{\hat \mu}}
({{g}E_C^{\epsilon}/\tau^{2-\epsilon}})
\ee
In the $M\to\infty$ limit,
the Lagrange multiplier enforcing the local normalization condition
$
{\bf w}^2_i(\tau)=1
$
tends to 
a spatially-and temporally-independent value $\lambda_i(\tau)=\lambda$ which can be found from the mean-field integral equation 
\be 
\sum_{\bf q}\int_{\omega} 
{1
\over {\omega^2/E_C({\bf q})
+z{g}E_C^{\epsilon}|\omega|^{1-\epsilon}+i\lambda}}
=1
\ee
where the propagator of the $\bf w$-field is read off from (19)
and $E_C({\bf q})$
is given by the Fourier transform
of the intra/inter-site capacitance matrix 
$C_{ij}$. 

In the FL case ($\epsilon=0$) the (real-valued) 
mean-field average $<i\lambda>$ 
remains finite for all values of the 
dimensionless parameter $g\sim (t/J)^2$, 
thus signalling the inescapable onset of
the classical CB with a reduced, yet finite, Coulomb gap:  
$E_C^{*}=<i\lambda>=E_C\exp({-O(zg)})
$
and 
$
E_C(1-O(zg))
$
for $zg>>1$ and $zg<<1$, 
respectively [42-49].

Qualitatively, this insulating behavior persists for all $\epsilon<0$ where the kernel (20) is super-Ohmic as, e.g., in the universal regime (12) which, if applicable, would formally correspond to $\epsilon=-1/2$. 

In contrast, for $0<\epsilon<1$ the integral (21) remains finite even in the limit of $\lambda\to 0$, thanks to the sub-Ohmic dissipative term. It then gives rise to a finite critical conductance 
\be 
{g}_c=1/(z\epsilon^{1+\epsilon})
\ee
above which $\lambda=0$, thereby signalling a quenching of CB 
and onset of a metallic behavior. In the ordinary FL granular materials such a behavior could only occur in the presence of sub-Ohmic dissipation due to, either, a coupling to external bath or intrinsic excitonic effects [42-45]. 

In terms of the critical
tunneling amplitude the transition occurs at
$
t_c\approx 
J^{1-\epsilon}E_C^{\epsilon}/
(z\epsilon^{1+\epsilon})^{1/2}
$
and its only dependence on the lattice structure is through $z$. 
Upon approaching the FL ($\epsilon\to 0$) the transition becomes unattainable. 

Also, in the customary case of $q=4$ such transition takes place at 
$t_c\sim (JE_C)^{1/2}$ (or, equivalently, $E_C\sim t^*$), in agreement with the earlier conclusions drawn for the SYK-lattices [21-36].  

Upon moving deeper into the insulating phase  
the renormalized Coulomb (Mott) gap rises as dictated by Eq.(21) 
\be
E_C^*= E_C(1-g/g_{c})^{\nu}/\epsilon^2
\ee
with the critical exponent $\nu=(1-\epsilon)/\epsilon$.

Notably, for $q=4$ the gap scales linearly with a deviation from the critical point
while for $q\to\infty$ the gap emerges abruptly and the transition resembles that of first order. 

\vspace{0.3cm}
\noindent
{\bf {\it Conductivity}} 
\vspace{0.3cm}

\noindent
The charge transport properties of a granular array can be  
assessed by computing the conductivity
\be
\sigma_{\mu\nu}(\omega)={ia^{2-d}\over \omega}
\int_{\tau}e^{i\omega\tau}
(\Pi_{\mu\nu}^{dia}(\tau)+\Pi_{\mu\nu}^{para}(\tau))
|_{\omega\to -i\omega+0^{+}}
\ee
where $d$ is the spatial dimension.

The dia- and para-magnetic contributions towards the overall 
conductivity read 
\bea
\Pi_{\mu\nu}^{dia}(\tau)
=
\delta_{\mu\nu}
{g\over \pi}
\int_{\tau^{\prime}}
(\delta(\tau)-\tau\delta(\tau-\tau^{\prime}))
\nonumber\\
\sum_{\rho}K(\tau^{\prime})
<\cos(\Phi_{i,i+\rho}(\tau)-\Phi_{i,i+\rho}(\tau^{\prime}))>
\eea
and
\bea
\Pi_{\mu\nu}^{para}(\tau)= 
{g\over \pi}\int_{\tau^{\prime},\tau^{\prime\prime}}
K(\tau-\tau^{\prime})K(\tau^{\prime\prime})~~~~~~~~~~~
\\
<\sin(\Phi_{i,i+\mu}(\tau)-\Phi_{i,i+\mu}(\tau^{\prime}))
\sin(\Phi_{i,i+\nu}(0)-\Phi_{i,i+\nu}
(\tau^{\prime\prime}))>
\nonumber
\eea
As in the Ohmic case [46-49] one can show that the dominant contribution 
comes from the $1^{st}$ order diamagnetic term while the corresponding $2^{nd}$ 
order correction cancels against the paramagnetic one. 

Besides, in contrast to the case of a single junction where the dominant (albeit subleading, $\sim g^2$) contribution towards the low-$T$ conductance is provided by inelastic co-tunneling processes [11-15], the latter appear to be suppressed exponentially 
with the size of the array [46-49].

Keeping the diamagnetic term one then arrives at the formula 
\bea
\sigma_{\mu\nu}(T)=a^{2-d}
\sum_{\bf q}
\int_{\omega}{1\over \omega} 
{\partial n(\omega)\over \partial\omega}
{\bf s}^{\mu}_q
{\bf s}^{\nu}_q
\nonumber\\
\int_{\tau} 
K(\tau)(1-\cos\omega\tau)
e^{-W(\tau)}
\eea
where
${\bf s}_{\bf q}=\partial_{\bf q}c_{\bf q}$
is a gradient of the sum over the nearest neighbors
$
c_{\bf q}=\sum_{\hat\mu}
(1-e^{i{\bf q}{\hat\mu}})
$, 
and the Debye-Waller (DW) 
which stems from 
the Gaussian averaging of the 
exponentials of the phase field
is given by the exponential of 
\be
W(\tau)=\sum_{\bf q}\int_{\omega} {\bf s_q}^2
(1-\cos\omega\tau)
<|\delta\Phi(\omega,{\bf q}|^2)>
\ee
Computing (27) one finds   
an approximate, yet practically 
convenient, expression for the longitudinal conductivity in terms of the Fourier transform ${\tilde K}(\omega)$
of the kernel (15) 
\be
\sigma(T)\sim e^{-W(1/2T)}{\tilde K}(T)/T
\ee
proposed 'ad hoc' in the early work of Refs.[42-45].

Away from criticality the phase fluctuations propagator entering the DW factor (28) reads 
\be
<|\delta\Phi(\omega,{\bf q}|^2>=
{1\over {\omega^2/E_C({\bf q})+gE_C^{2\epsilon}|\omega|^{1-2\epsilon}c_{\bf q}}}
\ee
In the FL case ($\epsilon=0$), the diffusion term ${\cal D}{\bf q}^2$ appearing in Eq.(11) derived by virtue of a phenomenological gradient expansion can be identified with (and absorbed into) that proportional to the conductance $g$, whereas for $\epsilon>0$ it can be neglected altogether 
as compared to the (singular) latter term. 

It is worth pointing out that the momentum sum in (28) turns out to be non-singular even in the potentially problematic dimensions $d=1$ or $2$, the only information about the lattice being its coordination number. 

In the deep CB regime corresponding to 
$g\ll g_c$ one might need to keep track 
of the large phase field fluctuations when computing the DW factor 
\bea
<e^{i\Phi_i(\tau)}e^{-i\Phi_i(0)}>
=e^{-{1\over 2}E^{*}_C\tau}
%\nonumber\\
\sum_ne^{-{E^*_C\over 2T}n(n+2T\tau)+2\pi n{\cal E}}~\nonumber\\
~~~
\eea
where the infinite sum over the winding numbers restores the periodicity
under $\tau\to\tau+1/T$.

The potential importance of the 
large phase fluctuations (hence, non-trivial winding numbers) brings about a conductance dependence on the offset charges $Q_i$. 
Their effects can be studied by restoring the topological '$\theta$-term' originating from the cross-terms in the charging energy (4),  
$
S_{top}=i\sum_i(Q_i-2\pi T{\cal E}/E_C)
\int_{\tau}\partial_{\tau}\Phi_i
$,
which accounts for the $T$-dependence of the intrinsic excess charge on the dots through the relation 
$2\pi{\cal E}=-\partial\mu/\partial T|_{T\to 0}$ [1-10]. 

In that regard, our discussion  
pertains to the charge quantization plateaus $Q_i=n$ where the (renormalized) Coulomb gap is maximal and the system is least likely to go metallic.
In contrast, at the transition points between the plateaus ($Q_i=n+1/2$) where the bare gap $E^*_C=E_C(1-2<Q_i>)$ vanishes, 
the conductivity takes its maximal values. The discussion of such (near)degenerate regime will appear elsewhere.

Nonetheless, at low $T$ the non-trivial winding numbers can be neglected 
and for $gE_C^*<T<E_C^*$
the conductivity governed by the 
$n=0$ term in the sum (31) 
shows the ordinary Arrhenius behavior
\be  
\sigma(T)\sim\sigma_0\exp(-E^*_C/T)
\ee
 where 
$\sigma_0\sim a^{d-2}g$.
For $\epsilon=0$ the insulating behavior sustains at all $g$.
 
As tunneling increases or temperature decreases, $T<gE_C^*$, Eq.(29) yields 
\be   
\sigma(T)\sim\sigma_0({T/gE_C^*})^{1/\pi zg} 
\ee
Expanding the DW factor to $1^{st}$ order 
reproduces the (negative) logarithmic  
(in all dimensions) conductivity correction, $\sigma(T)/\sigma_0=
1-O(1/zg)\ln(gE^*_C/T)$. 

Interestingly enough,
the above result appears to be accurate 
to the next, $2^{nd}$, order
due to the aforementioned cancellation between the higher order diamagnetic and paramagnetic corrections  [46-49]. 

As temperature decreases, the (negative) logarithmic conductivity correction gets cut off at energies $\sim g\delta$ (the rate of fermion escape from a dot) and becomes comparable to the bare conductivity
for $g\sim (1/z)\ln(E^*_C/\delta)$, again  
in agreement with the results of Refs.[46-49]. 

In contrast, for $\epsilon>0$ the conductivity suppression due to the DW factor remains non-singular at $T\to 0$ and Eq.(29) 
demonstrates the NFL power-law  
\be
\sigma(T)\sim
\sigma_0e^{-W(0)}{(E_C^*/T)^{2\epsilon}}
\ee
governed by a generically non-integer exponent. 

Incidentally, though, for $q=4$ Eq.(34) features a linear resistivity, consistent with the experimental data on a variety of the prospective SM compounds [37-41].

However, 
with increasing temperature the DW factor starts to contribute as well,  
resulting in a competition between the 
'kinematic' power-law (34) dictated by the SYK propagator (9)
and the fractional-exponential 
$T$-dependence of the correlation-induced 
$W(1/2T)$ 
\be
\ln\sigma(T)/\sigma_0= (T/E_C^*)^{2\epsilon}/(2\epsilon zg)-2\epsilon\ln{T/E_C^*}
\ee
The conductivity behavior switches from decreasing to growing, as signified by the  
sign change of $\ln\sigma(T)$,  
at $T=t^*=(2\epsilon z)^{1/2\epsilon}
t^{1/\epsilon}J^{1-1/\epsilon}$, 
consistent with the previously quoted value of $t^*$ for $q=4$ [21-36].

\vspace{0.3cm}
\noindent
{\bf {\it Density of states}} 
\vspace{0.3cm}

\noindent
Another important marker of the metal-insulator transition is a concomitant 'zero-bias anomaly' in the fermion density of states (DOS).
Evaluating the latter with the use of the factorization formula (13) one obtains
\bea
\nu(\omega)=
{1\over \pi}Im\int_{\tau} e^{i\omega\tau} 
{\cal G}(\tau)e^{-W^{\prime}(\tau)}|_{\omega\to -i\omega+0}
\eea
This time around the DW factor stands for the average
of only two (rather then four, as in Eq.(27)) exponentials of the phase field
\be
W^{\prime}(\tau)
= \sum_{\bf q}\int_{\omega}
(1-\cos\omega\tau)
<|\delta\Phi(\omega,{\bf q}|^2)>
\ee
Also, as opposed to the momentum sum in Eq.(28) its counterpart (37)  
appears to be rather sensitive 
to the spatial dimension,
which dependence is not limited 
to that on (and, in fact, does not involve)
the coordination number $z$.

In particular, for $d=2$ the momentum sum is logarithmic, thus reproducing 
 the log-normal 'zero-bias anomaly' familiar from the general 
theory of $2d$ disordered conductors [46-49]
in the FL case ($\epsilon=0$)
\be
\nu(\omega)\sim {1\over J}
\exp(-{1\over {\pi g}}\ln^2({gE^*_C/\omega}))
\ee 
The lack of information about the lattice 
in Eq.(38) can be understood from the fact that the momentum sum in (37) is dominated by 
small (rather than large, as in (28)) 
momenta.

By comparison, for a generic $\epsilon>0$ and $d=2$ one obtains the tunneling DOS 
\be
\nu(\omega)\sim J^{-1}
({\omega/J})^{(1/\epsilon g^{1/(1+\epsilon)})-\epsilon}
(J/t)^{O(1/\epsilon^2g)}
\ee
For $g>>1$ and $q=4$ Eq.(39) behaves as $1/\omega^{1/2}$, reproducing the salient $SYK_4$ transport feature [11-15]. 

The overall sign of this power-law dependence changes from negative (SM) to positive (MI) at the critical conductance 
$g_c^{\prime}=1/\epsilon^{2(1+\epsilon)}$ which appears to be 
generally consistent with (22).  

Instead, for $d\geq 3$ the momentum sum in (37) becomes non-singular, thus making the DW factor finite at all $\tau$ and resulting
in a generic linear DOS for $\omega\ll E_C^*$
\be
\nu(\omega)\sim {\omega/{J^{1-\epsilon}(E^*_C)^{1+\epsilon}}}
\ee
thus showing the development 
of a 'soft' gap.

The latter is  
markedly different from, both, the hard gap, 
$\nu(\omega)\sim\theta(\omega-E^*_C)$,  
which is a hallmark of the CB in a FL 
with momentum-dependent dispersion, as well as  
the bare DOS of the degenerate species, $\nu(\omega)\sim\delta(\omega-E^*_C)$.  

Thus, by measuring the tunneling DOS one might be able to access the properties of the physical fermion propagator across all the different regimes. Overall, its evolution with energy/temperature can be summarized as follows.

At $\omega, T\gg J$ it is that of free dispersionless fermions, $G(\tau)\sim sgn(\tau)$, which corresponds to the bare fermion dimension $[\psi]_0=0$ 
under the time dilation ($\tau\to l\tau$).
However, as the scale drops below $J$ and the system enters the SM regime
it evolves towards the SYK mean-field value 
$[\psi]_{SYK}=\Delta$.

Further, once the system cools down to $T\lesssim t^*$, the strongly relevant tunneling term continues 
to monotonically 
drive the dimension from the 
SYK value $\Delta$ towards the FL one,  
$[\psi]_{FL}=1/2$. 
Eventually, the SM gives way to a disordered FL if $g>g_c$ (or, equivalently,  $E_C^*<t^*$) , or a MI if $g<g_c$ (or $E_C^*>t^*$).

\vspace{0.3cm}
\noindent
{\bf {\it Discussion}} 
\vspace{0.3cm}

The above scenario of the MIT in a granular SYK array can be viewed as being somewhat complementary to that presented in the recent Ref.[56]. Rather than the CB effects, that work was mainly concerned with the effects of the Schwarzian fluctuations.

On the technical side, the renormalization group (RG) equations derived in Ref.[56] contain a conveniently chosen 
scale-dependent fermion dimension 
$[\psi](l)$. 
In the standard RG procedure, though, 
$[\psi]$ should have instead been found from the corresponding fermion field renormalization factor - which, in turn, would have to be computed as a sole function of the independently determined dimensionless RG charges obeying their own closed system of equations.
 
Besides, the MIT studied in Ref.[56]
occurs at the low tunneling strength, $t\sim J/N$, thus implying that for $N\gg 1$ and finite temperatures the system behaves as a metal for all the practical purposes. 

{ As compared to the above result, our analysis focuses on the role of the charging effects and predicts the onset of metallic behavior in the SYK array upon increasing the tunneling strength (or, equivalently, the inter-site conductance $g$) past the $N$-independent threshold value (see  Eq.(22)).

While being, at first sight, similar to the observations made in Refs.[21-36] our findings appear to be starkly different, as far as, both, the underlying mechanism and the actual critical parameter values are concerned. Besides, the proposed scenario  has no analogue in the case of the FL granular system without an additional source of sub-Ohmic dissipation. 

It should be noted, though, that the standard influence functional approach used in this (as well as much of the previous) work} is only applicable when all the relevant energy/temperature scales, such as $J, t^*, E_C^*$, etc. exceed  the average single-particle level 
spacing $\delta\sim J/N$ (moreover, its many body counterpart, $\delta_{N}\sim J\exp({-O(N)})$ [1-3,50-51]). 
 
Incidentally, at energies of order $\delta$ the renormalizing effects of the Schwarzian fluctuations would have just started to develop and the universal regime (12) {(let alone the MIT scenario of Ref.[56])} could not have yet been reached. 

Furthermore, as recently shown in the case of a single tunnel junction [15], at such low energies one might expect an intricate competition between the SYK, charging, tunneling, as well as 
single- and multi-level Kondo phenomena. 
Therefore, for a complete picture it might be necessary to consider the charging and SYK effects on equal footing with the potentially important local Kondo resonances. 

Lastly, by assuming the simplest nearest-neighbor tunneling, we deliberately left out such subtle topics as variable range hopping (Mott, Efros-Shklovskii, and related   
mechanisms, all capable of yielding $\sigma(T)\sim\exp(-(E_C^*/T)^{\nu})$ with various fractional exponents $\nu$) whose inclusion is likely to be necessary if a detailed comparison with the experimental data on actual SYK arrays were ever to be made. 

{Considering the long and still unfinished history of  studies of the phenomenon of CB even in the ordinary FL granular materials  
it would be rather unrealistic to try to cover a potentially rich variety of the pertinent regimes all at once. One might hope, however, that the present attempt to shed some light on the new aspects of this long-standing problem could revitalize the field as a whole.}

\vspace{0.3cm}
\noindent
{\bf {\it Acknowledgements}} 
\vspace{0.3cm}

\noindent
The author acknowledges hospitality at 
and support from the Pauli Center for Theoretical Studies (ETH, Zurich).

%\pagebreak

%\end{thebibliography}
\end{document}